\documentclass[reprint,amsmath,amssymb,aps,
]{revtex4-1}

\usepackage{graphicx}
\usepackage{dcolumn}
\usepackage{bm,epstopdf}

\begin{document}

\title{General consensus with circular opinion under attractive and
repulsive mechanisms}

\author{Shun Gao}
 \affiliation{College of Physics and Electronic Engineering, Sichuan Normal University, Chengdu, 610101, People's Republic of China}
\author{Wenchen Han}
\email{wchan@sicnu.edu.cn}
\affiliation{College of Physics and Electronic Engineering, Sichuan Normal University, Chengdu, 610101, People's Republic of China}
\author{Changwei Huang} \affiliation{School of Computer, Electronics and Information, Guangxi University, Nanning, 530004, People¡¯s Republic of China}
\author{Junzhong Yang}
\affiliation{School of Science, Beijing University of Posts and Telecommunications, Beijing, 100876, People's Republic of China}


\begin{abstract}
In this work, we study a nonlocal opinion dynamics in a ring of agents with
circular opinion in the presence of both attractive and repulsive interactions.
We identified three types of consensus in this model, including global
consensus, local consensus, and chimera consensus. In global consensus,
both local agreement among adjacent agents and global agreement among all
agents are achieved. In local consensus, local agreement is satisfied but
global agreement fails. There are two domains in chimera consensus,
one preserves local agreement and the other breaks the local agreement.
The relation between the opinion difference between adjacent agents and
the interaction radius is investigated and a scaling law is found. The
transitions between local consensus and chimera consensus are exemplified.
\end{abstract}

\maketitle


\section{Introduction}

Everyday, we encounter a variety of situations in which we have to make
decisions following our opinions, typical examples including political
campaigns \cite{fri16,gal17}. Human society could be free of conflicts if
all agents share the same opinion on issues they encounter. On the other hand,
human society may be more energetic if different opinions coexist with each other.
Opinion dynamics models the opinion formation by focusing on the interaction and
the communication among individuals. There exist different types of models on
opinion dynamics. Opinion model may be classified into discrete opinion models
and continuous opinion models by whether opinion is represented as discrete
numbers or continuous ones. The voter model \cite{dur12}, the Galam
majority-rule model \cite{gal02}, and the Sznajd model \cite{szn00} are
examples for discrete opinion models. Degroot model \cite{deg74},
Friedkin-Johnsen model \cite{fri90}, Deffuant-Weisbuch (DW) model \cite{deff00},
and Hegselmann-Krause model \cite{heg02} are typical continuous opinion models.

The DW model is the one of the most famous continuous opinion models. In the
DW model, there exists a bounded confidence. Two agents interact with each other
by reducing their opinion difference with a convergence rate if their opinion
difference is less than the bounded confidence. Depending on the bounded
confidence, the model displays different asymptotic states, fragmentation where
several opinion clusters coexist, polarization where there only exist two opinion
clusters, and consensus where all agents share a same opinion. The concept of
consensus is widely applied in different disciplines, such as
sociophysics \cite{gal13,baro18}, management \cite{gong15,gong15b}, and
automatical control \cite{olf07,gong17}, which has drawn much attention on it.
Concerning with the consensus, many mechanisms have been proposed, including
the introduction of inflexible minorities \cite{gal07,col16}, heterogeneous
bounded confidence \cite{lor07,han19}, heterogeneous convergence
rate \cite{huang18}, effects of social power \cite{jal13},
effects of leadership \cite{dong17}, evolutionary games based
model \cite{yang16}, and the effects of communication burstiness \cite{doy17}.

The DW model is actually defined on an all-to-all network in which each agent
may interact with the rest of population provided that the opinion difference
between them is within the bounded confidence. In reality, interaction among
agents always forms sparse complex networks. Recently, opinion dynamics on
sparse complex networks has been investigated where agents can interact with
their nearest neighbors \cite{lee16,yu10}. Besides all-to-all networks
accounting for global interaction among agents and sparse networks for local
interaction, nonlocal interaction where every agent interacts with a finite
fraction of population also draws interests from scientists. Nonlocal interaction
may induce interesting dynamics. It has been shown that nonlocally coupled
oscillators may produce a intriguing dynamical state, chimera state, in which
coherent domains coexist with incoherent ones \cite{abra04}. It is interesting
to investigate whether nonlocal interaction can induce interesting dynamics
in opinion models. Then, the DW model considers the attractive interaction
in which agents tend to reduce their opinion differences. However,
repulsive interaction is also common in reality, which has been modeled
in many social systems \cite{rad09,mar10,huet08}. In opinion models,
the role of repulsive interaction has been studied by assigning links among
agents to be either attractive ones or repulsive ones. However, consider
two agents. It is more likely for them to be enemy if their opinions are
far away from each other and to be friend otherwise. In modelling opinion
dynamics, friends tend to adopt attractive interaction while enemies tend
to adopt repulsive interaction \cite{chen17}. In most of continuous opinion
models, opinions are represented as real numbers in the range from $0$ to $1$.
The description of opinion like this allows for the existence of extreme opinions,
for example the opinion $0$ and the opinion $1$, and its simplification goes to
binary opinion. There are some works considering circular opinion where opinions
are represented as real numbers on a ring with unit length and some interesting
phenomena have been found \cite{chen17}. In the description of circular opinion,
the extreme opinions are absent. The circular opinion could be justified by social
phenomenon that people convert their faith from one religion to
another~\cite{leman10}.

In this work, we investigate a nonlocal DW model with circular opinion under
attractive and repulsive interactions on a ring-like network. We identify three
types of consensus, global consensuses, local consensus, and chimera consensus.
All agents share a same opinion in the global consensus. In local consensus,
the opinions of adjacent agents are close to each other while the opinion
difference between distant agents are sufficiently large. In other words,
local agreement among agents exists while global agreement is absent in local
consensus. Chimera consensus is a new type of state involving consensus
in which local agreement is violated within a domain of agents and is preserved
in rest of population.

The rest of this paper is organized as follows. In section \ref{sect2}, we
introduce the model. The numerical results and discussions are presented in
section \ref{sect3}. The properties of three types of consensus and their
dependence on the interaction range of agents are investigated. Finally,
the conclusion is drawn in section \ref{sect4}.

\section{Model} \label{sect2}
We consider a population of $N$ agents sitting on a ring. Each agent is
represented by a node index $i$, which is taken module $N$. We assume nonlocal
interaction among agents in which every agent may interact with $k$ neighbors
on each side. That is, agents $i$ and $j$ may interact with each other if
$\min\{|i-j|,N-|i-j|\}\leq k$. The interaction among agents reduces to a local
one for $k=1$, while it becomes a global one when $2k=N-1$. For convenience,
we define the interaction radius $p=k/N$, which is in the range $(0,0.5)$.
The opinion of agent $i$ is represented as $x_i(t)\in[0,1)$ at the time step $t$.
We consider circular opinion $x_i\in[0,1)$ such that $x_i(t)=(x_i(t) \mod 1)$.
For circular opinion, the opinion difference between agents $i$ and $j$ is defined
as $|\delta_{i,j}|=\min\{|x_i-x_j|,1-|x_i-x_j|\}$. Therefore, the upper bound of
opinion difference between agents is $0.5$. The updating rule of agents' opinions
follows a modified Deffuant-Weishbuch rule incorporating both attractive and repulsive interactions \cite{chen17}, which takes into considerations a pairwise
interaction one. There are two parameters, the bounded confidence
$\sigma\in(0,0.5]$ and the convergence rater $\mu\in(0,0.5]$. Initially,
each agent is assigned an opinion randomly chosen from the interval $[0,1)$
or specified. At each time step, two neighboring agents, agent $i$ and agent $j$,
are chosen at random. If the opinion difference between them is less than the
bounded confidence ($|\delta_{i,j}(t)|<\sigma$), they adopt attractive
interaction and update their opinions to get closer. If their opinion difference
is larger than the bounded confidence ($|\delta_{i,j}(t)|>\sigma$), they adopt
the repulsive interaction to drive their opinions further away from each other.
Combining the attractive and repulsive interactions together, the opinion
updating rule is written as
\begin{equation} \label{eq1}
\left\{
\begin{aligned}
x_i(t+1)=x_i(t)+\mu\Theta(\sigma-|\delta_{i,j}(t)|)\delta_{j,i}(t),\\
x_j(t+1)=x_j(t)+\mu\Theta(\sigma-|\delta_{i,j}(t)|)\delta_{i,j}(t).\\
\end{aligned}
\right.
\end{equation}
where $\Theta(y)$ is defined as $\Theta(y)=1$ for $y>0$ and $\Theta(y)=-1$
otherwise. One Monte Carlo time step consists $N$ such events. Throughout
this work, the opinions of agents are updated asynchronously.

We characterize the asymptotic states in the model using two methods. Firstly,
we observe the opinion profile displaying the agents' opinions $x_i(t)$ against
their locations. Secondly, we monitor the average opinion difference between
adjacent agents, which is defined as
\begin{equation} \label{eq2}
\Delta=\langle\sum\nolimits_{i=1}^{N}{|\delta_{i+1,i}|/N}\rangle_t
\end{equation}
where $\langle\cdot\rangle_t$ means the time average over $100$ Monte Carlo
time steps after $10^6$ transient time steps.

\section{Results and discussion} \label{sect3}

\begin{figure}
\includegraphics[width=3.4in]{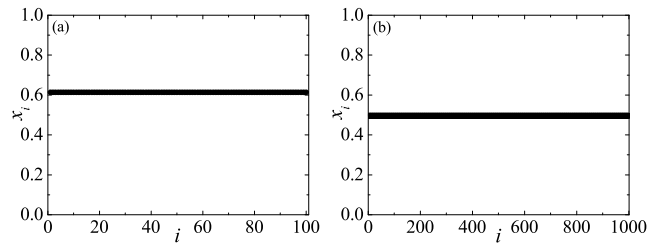}
\caption{\label{fig_1} Opinion profiles for global consensus ($GC$). (a) $N=100$
and $p=0.05$. (b) $N=1000$ and $p=0.05$. The other parameters are fixed with the
bounded confidence $\sigma=0.3$ and the convergence rate $\mu=0.2$.}
\end{figure}

We focus on consensus. The bounded confidence determines the asymptotic state
in opinion model and $\delta>0.5$ assures the realization of consensus in typical
DW models. The convergence rate $\mu$ is always set to be $0.5$ in most of works.
However, it has been shown that $\mu$ plays important role in opinion dynamics
by controlling the time scale in the evolution of opinion. Considering the
existence of both the attractive and repulsive interactions among agents,
we set $\mu=0.2$ to foster the competition between these two types of interaction.
Furthermore, we set $\sigma=0.3$ for the purpose to realize consensus. The state
of consensus is characterized by the opinion profile and the average opinion
difference between adjacent agents $\Delta$. Especially, we distinguish three
types of consensuses through the opinion profile.

The first type of consensus is the global consensus ($GC$) which refers to a
state where all agents hold a same opinion (the global agreement) and is the
one found in typical DW models. Its opinion profile is a straight horizontal line,
as shown in Fig.\ref{fig_1}. The opinion difference between adjacent agents is
$0$ in the time limit $t\rightarrow\infty$. However, different from typical DW
models \cite{jac06,shang13}, the average opinion of the initial opinion may not
be the final opinion of all agents, due to the circular opinion mechanism.
In $GC$, only the attractive interaction takes effect. To be noted, $GC$ is
always stable in the model, which is independent of the system size $N$ and
the interaction radius $p$. Perturbation to GC dies off quickly due to the
attractive interaction.

\begin{figure}
\includegraphics[width=3.4in]{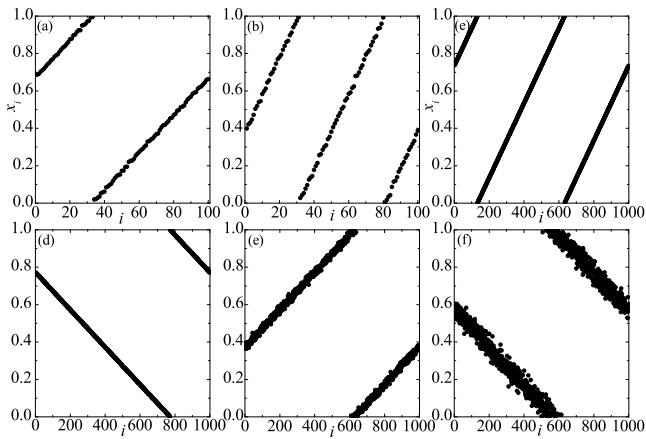}
\caption{\label{fig_2} Opinion profiles for local consensus ($PC$). (a) $PC_1$
at $N=100$ and $p=0.01$. (b) $PC_2$ at $N=100$ and $p=0.01$. (c) $PC_2$ at
$N=1000$ and $p=0.001$. (d) $PC_1$ at $N=1000$ and $p=0.005$. (e) $PC_1$ at
$N=1000$ and $p=0.05$. (f) $PC_1$ at $N=1000$ and $p=0.15$. The other parameters
are fixed with the bounded confidence $\sigma=0.3$ and the convergence rate
$\mu=0.2$.}
\end{figure}

The second type of consensus is the local consensus ($LC$). Figure~\ref{fig_2}
shows the opinion profiles for $LC$ at different parameters. In a typical $LC$,
the opinion makes $n$ full turns around the opinion space when it traverses from
agent $1$ to agent $N$ and, at the same time, $x_i$ changes with $i$ monotonically.
We denote $LC$ with $n$ full turns of opinion as $LC_n$. The local agreement
refers to the situation in which the opinion differences between any adjacent
agents are sufficiently small. In an $LC$, the local agreement among adjacent
agents is achieved but the large opinion difference between distant agents
breaks the global agreement. Figures~\ref{fig_2} (a-c) show $LC$s for $p=1/N$
(the situation with local interaction). Clearly, $LC_n$ with different $n$ may
be realized depending on initial condition and the realization of $LC_n$ is
independent of the system size $N$. Then, we present $LC$s in Figs.~\ref{fig_2}
(d-f) for nonlocal interaction. We find that $x_i$ still changes with $i$
monotonically but in the sense of an overall trend. In comparison with the
situation with local interaction, the exact order of $x_i$ with $i$ is lost since
agents with a little large distance may still be neighboring ones and they may
interact with each other. As a result, the opinion difference between adjacent
agents fluctuates around zero and the fluctuation increases with $p$. These
features reflect the characteristics of $LC$s. In $LC$s, neighboring agents tend
to reach local agreement among them. For large $p$, every agent is required
to be in local agreement with his neighbors on both sides. This trades off
the loss of exact order of $x_i$ with $i$. On the other hand, the attractive
interaction among the neighboring agents assures the opinion convergence
locally even though the opinion differences among them may be large. Thus,
the opinion profiles of $LC$s for large $p$ display fluctuations and the
monotonic variation with $i$ in an overall way. To be mentioned, since the
local agreement exists in $LC$s,
attractive interaction among agents still plays dominant role.

\begin{figure}
\includegraphics[width=3.4in]{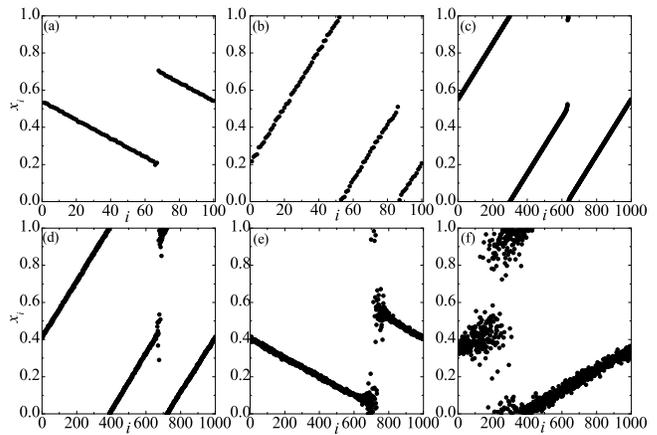}
\caption{\label{fig_3} Opinion profiles for chimera consensus ($CC$). (a)
$CC_1$ $N=100$ and $p=0.01$. (b) $CC_2$ at $N=100$ and $p=0.01$. (c) $CC_2$
at $N=1000$ and $p=0.001$. (d) $CC_2$ at $N=1000$ and $p=0.01$. (e) $CC_1$
at $N=1000$ and $p=0.05$. (f) $CC_1$ at $N=1000$ and $p=0.15$. The other
parameters are fixed with the bounded confidence $\sigma=0.3$ and the
convergence rate $\mu=0.2$.}
\end{figure}

$GC$s result from attractive interaction while circular opinion, together with
attractive interaction, are responsible for $LC$. When the effect of repulsive
interaction is involved, we find the third type of consensus, chimera consensus
(CC). Typical $CC$s are shown in Fig.~\ref{fig_3}. Figures~\ref{fig_3} (a-c)
show the results for local interaction with $p=1/N$. The opinion $x_i$ makes
a half turn when agent goes from 1 to $N$ £¨denoted as $CC_1$ in
Fig.~\ref{fig_3} (a) while one and a half turns (denoted as $CC_2$ in
Figs.~\ref{fig_3} (b,c). The local agreement is violated only for one pair of
agents whose opinion difference becomes $\sigma$, which implies the repulsive
interaction between them. Figures~\ref{fig_3} (d-f) for nonlocal interaction
show that $x_i$ also performs an extra half turn after several full turns when
$i$ goes through $1$ to $N$ and the local agreement fails in a finite domain.
In the domain breaking the local agreement, agents' opinions have a gap around $0.5$,
which is allowed by the repulsive interaction, and they fluctuate
greatly from one agent to another, which is due to the nonlocal interaction.
The size of the domain breaking the local agreement is determined by the
interaction radius $p$ and displays a positive correlation with $p$.
The coexistence of domains supporting the local agreement and breaking the
local agreements in $CC$s resembles the dynamical chimera states in nonlocally
coupled systems \cite{omel11} if we treat the domain supporting (or breaking)
the local agreement as coherent (or incoherent) domain. In addition, in a $CC$,
agents' opinions in the domain holding the local agreement are frozen while
they fluctuate in time in the domain breaking the local agreement. Briefly,
$CC$s are much similar to $LC$s except that the local agreement is broken in
the incoherent domain.

\begin{figure}
\includegraphics[width=3.4in]{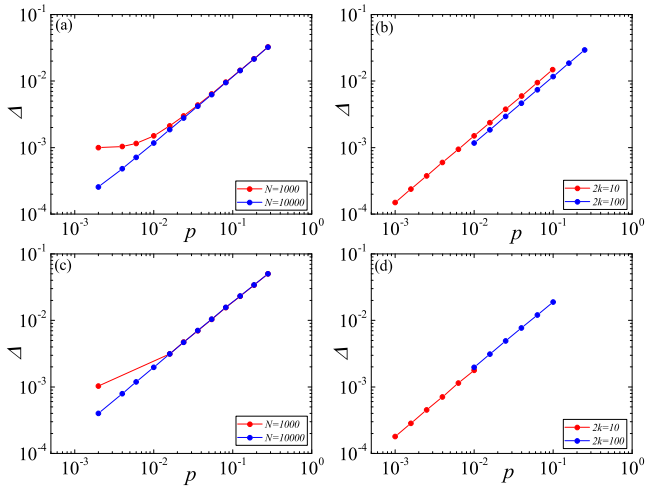}
\caption{\label{fig_4} The average opinion difference $\Delta$ against the
interaction radius $p$. (a) and (b) for $LC_1$, (c) and (d) for $CC_1$.
(a) and (c) for the fixed population size $N$, (b) and (d) for the fixed number
of neighbors of every agent $2k$. The other parameters are fixed with the
bounded confidence $\sigma=0.3$ and the convergence rate $\mu=0.2$.}
\end{figure}

In $GC$s, all agents share a same opinion, which means that the average opinion
difference $\Delta$ in the population is zero. However, the global agreement
is absent in $LC$s and $CC$s, which allows for nonzero average opinion difference
$\Delta$. Actually, $\Delta$ measures the fluctuation of the opinion differences
between adjacent agents. The above results have suggested that the fluctuation
of the opinion differences between adjacent agents strongly depends on the
interaction radius $p$ for $LC$s and $CC$s. Here, we consider the dependence
of $\Delta$ on $p$. Since there are plenty of $LC$s and $CC$s with opinion $x_i$
making different full turns, we just consider $LC_1$ with one full turn and $CC_1$
with a half turn for convenience. Including the opinion gap in the domain breaking
the local agreement in $CC$, opinion in the population transverses a full circle
in the opinion space for both $LC_1$ and $CC_1$. For each consensus state, we
consider two situations, one with fixed population size $N$ and the other with
fixed number of neighbors of every agent $2k$. As shown in Fig.~\ref{fig_4}, the
average opinion difference $\Delta$ always scales with $p$ at a same exponent
around $1$, which is independent of the type of consensus, the population size,
and the number of neighbors of an agent. The deviation of the dependence of
$\Delta$ on $p$ from the scaling law in Figs.~\ref{fig_4} (a,c) for $N=1000$
is induced by the fact that $\Delta$ is restricted by its minimum $1/N$.

\begin{figure}
\includegraphics[width=3.4in]{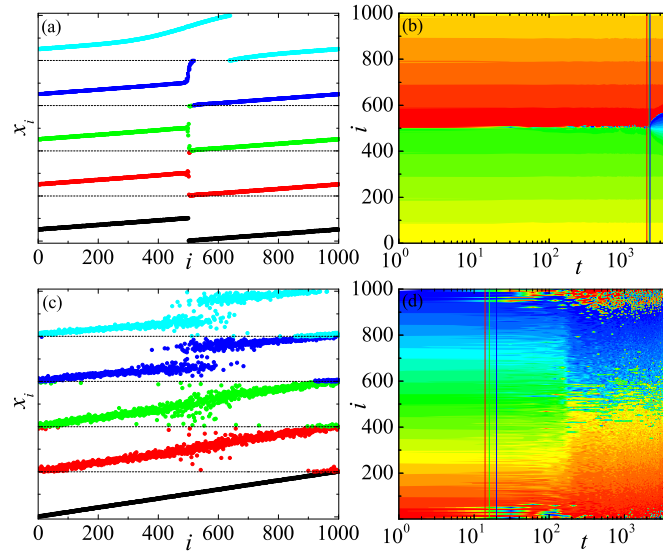}
\caption{\label{fig_5} The transitions between $CC_1$ and $LC_1$. (a) and (b)
show the evolution from $CC_1$ to $LC_1$ at $N=1000$ and $p=0.005$. (c) and (d)
show the evolution from $LC_1$ to $CC_1$ at $N=1000$ and $p=0.18$. (a) and (c)
show opinion profiles at different times from $t=0$ (black) to $t=10000$ (cyan).
The opinion profiles colored by red, green, and blue correspond to the vertical
lines in (b) and (d), respectively. (b) and (d) show the spatiotemporal evolution
of agents' opinions. The other parameters are fixed with the bounded confidence
$\sigma=0.3$ and the convergence parameter $\mu=0.2$.}
\end{figure}

It is of interest to have some discussions on the stabilities of these three
types of consensuses. First of all, our model allows for multistability and
these three consensuses may coexist in a large range of the interaction radius
$p$. Especially, $GC$ is always locally stable to weak perturbation. On the
other hand, $CC$s are more stable than $LC$s at large $p$ while $LC$s are more
stable than $CC$ at small $p$, which can be hinted in Fig.~\ref{fig_4}
(for example, there is no data for CCs at small $p$ and $N=1000$ in
Fig.~\ref{fig_4} (c). When $p$ is sufficiently high, both $LC$s and $CC$s yield
to $GC$. We take two examples with the population size $N=1000$ to show the
transitions between $LC$s and $CC$s. At $p=0.005$, we prepare a $CC_1$ as initial
conditions in which the opinion gap occurs between agents $i=500$ and $j=501$
and run the model.
Figure~\ref{fig_5} (a) shows five snapshots of the opinion profile at different
times and Fig.~\ref{fig_5} (b) shows the spatiotemporal evolution of agents'
opinions over $10^4$ Monte Carlo time steps. To illustrate how such a transition
occurs, we consider two agents locating on the two sides of the opinion gap,
agent $i_1$ (e.g., $i_1=498$) and agent $i_2$ (e.g., $i_2=502$). As time evolves,
$x_{i_1}$ (or $x_{i_2}$) could be pushed up (or pulled down) by his neighbors on
the right (or left) side of the gap through the repulsive interaction.
When $x_{i_1}-x_{i_2}$ becomes less than $\sigma$, the attractive interaction
between agents $i_1$ and $i_2$ steps in and draw them closer. Similar behaviors
may occur to other agents near the opinion gap. Resultantly, a $CC_1$ transforms
to a $LC_1$. To be noted, when $p=0.001$ (the local interaction), $CC$ is stable
since the absence of nonlocal interaction. At $p=0.18$, Figs.~\ref{fig_5} (c,d)
show the transition from a $PC_1$ to a $CC_1$. In this example, starting from
a $PC_1$, the local agreement is quickly broken in a certain area and the opinion
profile evolves into a $CC_1$.

In closing, we make some discussions. The three consensuses, $GC$, $LC$, and
$CC$, are differentiated by the breaking or not of the local agreement and
the global agreement. However, the local agreement always exists if we ignore
the opinion gap in the incoherent domain in $CC$s. Following this line, the
three types of consensuses may be termed under the same rule according to how
many turns opinion $x_i$ from agent $1$ to agent $N$ has made in the opinion
space. For example, $GC$ is actually a special case of $LC$ and can be denoted
as $LC_0$. $CC$ can be denoted as $LC_{\frac{n}{2}}$ with $n$ an odd integer.
Furthermore, we assume that positive $n$ suggests an upward trend in opinion
profiles while negative $n$ a downward trend. In this way, all of these
consensuses can be represented as $LC_{\frac{n}{2}}$ with integer $n$.

\section{Conclusion}   \label{sect4}

To conclude, we have studied the opinion dynamics in a ring of agents with
circular opinion under nonlocal attractive and repulsive interaction. Three
types of consensuses, including the global consensus, the local consensus, and
the chimera consensus, are identified by the opinion profiles. The global
consensus with the global agreement has been well studied in previous works.
The local consensus which only requires the local agreement was found only very
recently \cite{chen17}. The chimera consensus divides the population into two
domains, one preserves the local agreement and the other breaks the local
agreement. We studied the dependence of the average opinion difference between
adjacent agents on the interaction radius and found a scaling law between them.
We found that chimera consensus is more stable than local consensus at large
interaction radius while local consensus is more stable than chimera state at
small interaction radius.

\section{Acknowledgments}

This work is supported by the National Natural Science Foundation of China
under Grant Nos. $11575036$.

\end{document}